\documentclass[pra,twocolumn,superscriptaddress,showpacs,floatfix]{revtex4-1}
\usepackage{graphicx,amssymb}
\usepackage[fleqn]{amsmath}
\usepackage{amsfonts}
\usepackage{dcolumn}
\usepackage{bm}
\usepackage{braket}
\usepackage{multirow} 

\newcommand{\gras}[1]{\boldsymbol{#1}}

\DeclareMathOperator{\erf}{erf}

\begin{document}

\preprint{ }

\title{Bound and resonance states of the dipolar anion of hydrogen cyanide: competition between threshold effects and rotation in an open quantum system}

\author{K. Fossez}
\affiliation{Grand Acc\'el\'erateur National d'Ions Lourds (GANIL), CEA/DSM - CNRS/IN2P3,
BP 55027, F-14076 Caen Cedex, France}

\author{N. Michel}
\affiliation{Grand Acc\'el\'erateur National d'Ions Lourds (GANIL), CEA/DSM - CNRS/IN2P3,
BP 55027, F-14076 Caen Cedex, France}

\author{W. Nazarewicz}
\affiliation{Department of Physics and Astronomy and NSCL/FRIB Laboratory,
Michigan State University, East Lansing, Michigan  48824, USA}
\affiliation{Physics Division, Oak Ridge National Laboratory, P. O. Box 2008, Oak Ridge, Tennessee 37831, USA}
\affiliation{Institute of Theoretical Physics, Faculty of Physics, University of Warsaw, ul. Ho$\dot{z}$a 69, PL-00-681 Warsaw, Poland}

\author{M. P{\l}oszajczak}
\affiliation{Grand Acc\'el\'erateur National d'Ions Lourds (GANIL), CEA/DSM - CNRS/IN2P3,
BP 55027, F-14076 Caen Cedex, France}

\author{Y. Jaganathen}
\affiliation{Department of Physics \&
Astronomy, University of Tennessee, Knoxville, Tennessee 37996, USA}
\affiliation{NSCL/FRIB Laboratory,
Michigan State University, East Lansing, Michigan  48824, USA}

\date{\today}

\begin{abstract}
	{
		Bound and resonance states of the dipole-bound anion of hydrogen cyanide HCN$^-$ are studied using a non-adiabatic pseudopotential method and the  Berggren expansion technique involving bound states, decaying resonant states, and non-resonant scattering continuum. We devise an algorithm to identify the resonant states  in the complex energy plane.
		To characterize spatial distributions of electronic wave functions, we introduce the body-fixed density and use it to assign families of resonant states into collective rotational bands. We find that the non-adiabatic coupling  of electronic motion to molecular  rotation  results in a transition from  the strong-coupling to weak-coupling regime. In the strong coupling limit, the electron  moving in a  subthreshold, spatially  extended halo state follows the rotational motion of the molecule. Above the   ionization threshold, electron's motion in a resonance state becomes largely decoupled from molecular rotation.  Widths of resonance-band members depend primarily on the electron orbital angular momentum.
	}
\end{abstract}


\maketitle

\section{Introduction}

Dipolar anions are one of the most spectacular examples of
marginally bound quantum systems \cite{garrett70,*garrett71,Wong74,Jordan77,jordan03,desfrancois96,*abdoul98,Compt01,Desfrancois04,adamowicz85,gutsev97_125,gutsev98,Simons08}. Wave functions of electrons
coupled to neutral dipole molecules  \cite{Fermi47,levy67} are extremely extended; they form the extreme quantum halo states \cite{Rii2000,jens04,Mitr05,Knoop09,Hammer10,Ferlaino}.
 Resonance energies of dipolar anions, including those associated with rotational threshold states,  can been
determined  in  high resolution electron photodetachment experiments
 \cite{lykke84,Marks86,Andersen91,brinkman93,mullin93,ard09}. Theoretically,
 however,
 the literature on the unbound part of the spectrum of dipole potentials, and  multipolar anions in particular, is fairly limited \cite{Omalley65,estrada84,Clark84,Fabrikant85,clary88,clary89,McCartney90,sadeghpour00,Martorell08}.
 
The breakdown of the adiabatic approximation in  dipolar molecules possessing a supercritical moment \cite{garrett82,garrett10,camblong01_845,coon02_846,*bawin04_844,*bawin07_39,fossez13} caused by  coupling
of electron's motion to the rotational motion of the molecule,
is expected to profoundly impact the properties of rotational bands in such systems \cite{clary88,clary89,ard09,garrett10}, such as the
the number of rotationally excited bound anion states. 

In this study, we address the nature of the unbound part of the spectrum of dipolar anions. In particular, we are interested in elucidating the transition 
from the  rotational  motion of weakly-bound subthreshold states to the rotational-like behavior exhibited by unbound resonances. 
The competition between continuum effects, collective rotation, and non-adiabatic aspects of the problem makes the description of threshold 
states in dipole-bound molecules both interesting and challenging.

Our theoretical framework is based on the Bergggren expansion method (BEM) -- a complex-energy 
resonant state expansion \cite{berggren82,Beggren1993,lind1993} based on a completeness relation introduced by Berggren \cite{berggren1968} that involves bound, decaying, and scattering states. In the context of coupled-channel method, BEM was successfully applied to
molecules \cite{fossez13} and nuclei \cite{Fer97,Kruppa00,Barmore00,Kru04,Jag12,*Jag14,IdBetan08,*IdBetan14}. 
The advantage of this method, which is of particular importance to the problem of dipole-bound anions when the rotational motion of the molecule is considered \cite{Chernov05,fossez13}, is that  the BEM is largely independent of the precise implementation of  boundary conditions at infinity. This is not the case for other techniques such the direct  method of integrating coupled-channel equations.

The calculations have been carried out for the rotational spectrum
of dipole-bound anions of hydrogen cyanide HCN$^-$, which
has long served as a prototype  of a dipole-bound anion \cite{Klahn98,jordan03} and
was a subject of experimental and theoretical studies \cite{skurski01,peterson02,ard09}.
Here, we extend our previous studies \cite{fossez13} of bound states of dipolar molecules to the unbound part of the spectrum. 
To integrate coupled channel equations, we use the Berggren expansion method as it offers superior accuracy as compared 
to the direct integration approach for weakly-bound states and, contrary to direct integration approach, allows to describe unbound resonant states.

This paper is organized as follows. The model Hamiltonian is  discussed in Sec.~\ref{S:pseudopotential}. The coupled channel  
formulation of the Schr\"odinger equation for dipole-bound anions is outlined in Sec.~\ref{S:CC}.
The Berggren expansion method is introduced in Sec.~\ref{Berggren_basis_diag}. The parameters of our calculation 
are given in Sec.~\ref{parameters}. Section~\ref{s:reson} presents the technique adopted to identify the decaying Gamow 
states (resonances). To visualize valence electron distributions, in Sec.~\ref{s:density} we introduce the
intrinsic one-body density. The predictions  for bound states and resonances of  HCN$^-$ are collected in Sec.~\ref{results}.
Finally, Sec.~\ref{conclusions} contains the conclusions and outlook.
 
\section{Hamiltonian}
	\label{S:pseudopotential}
	
The dipolar anions are composed of a neutral polar molecule with a dipole moment ${ \mu }$ that is large enough to bind an additional electron. In the present study, the HCN$^-$ dipolar anion is described in the Born-Oppenheimer approximation, and the intrinsic spin of an external electron is neglected \cite{garrett82}, largely simplifying the equations \cite{garrett10}. Within the pseudo-potential method, the Hamiltonian of a dipolar anion can be  written as:
	\begin{equation}
		H = \frac{ \gras{p}_{e}^{2} }{ 2 {m}_{e} } + \frac{ \gras{j}^{2} }{ 2I } + {V}
		\label{H_Born_Opp}
	\end{equation}
	where ${ I }$ is the moment of inertia of the molecule, ${ \gras{p}_{e} }$ is the linear momentum of the valence electron, and ${ {m}_{e} }$ its mass.
	The electron-molecule interaction $V$ is approximated by a one-body pseudo-potential \cite{garrett79,garrett80,garrett82}:
	\begin{equation}
		V ( r , \theta ) = {V}_{ \text{dip} } ( r , \theta ) + {V}_{ \alpha } ( r , \theta ) + {V}_{ {Q}_{zz} } ( r , \theta ) + {V}_{ \rm SR } (r),
		\label{eq_main_pot}
	\end{equation}
	where ${ \theta }$ is the angle between the dipolar charge separation ${ \gras{s} }$ and electron coordinate;
	\begin{equation}
		{V}_{ \text{dip} } ( r , \theta ) = - \mu e \sum_{ \lambda = 1, 3, \cdots }
		{ \left( \frac{ {r}_{<} }{ {r}_{>} } \right) }^{ \lambda } \frac{1}{ s {r}_{>} } {P}_{ \lambda }( \cos\theta )
		\label{V_dip}
	\end{equation}
	is the electric dipole potential of the molecule; 
	\begin{equation}
		{V}_{\alpha} ( r , \theta ) = - \frac{ {e}^{2} }{ 2 {r}^{4} } \left[ { \alpha }_{0} + { \alpha }_{2} {P}_{2}( \cos\theta ) \right] f(r)
		\label{Valpha}
	\end{equation}
	is the induced dipole potential, where  ${ { \alpha }_{0} }$ and ${ { \alpha }_{2} }$ are the spherical and quadrupole polarizabilities of the linear molecule;
	\begin{equation}
		{V}_{ {Q}_{zz} } ( r , \theta ) = - \frac{e}{ {r}^{3} } {Q}_{zz} {P}_{2}( \cos\theta ) f(r)
		\label{VQzz}
	\end{equation}
	is the potential due to the permanent quadrupole moment of the molecule, and 
	\begin{equation}
		{V}_{ \rm SR } (r) = {V}_{0} \exp \left[ - { ( r / {r}_{c} ) }^{6} \right],
		\label{VSR}
	\end{equation}
	is the short-range potential, where ${ {r}_{c} }$ is a radius range.
	The short-range potential accounts for the exchange effects and compensates for spurious effects induced by the regularization function
	\begin{equation}
		f (r) = 1 - \exp \left[ -  ( r / {r}_{0} )^6 \right]
		\label{cutoff}
	\end{equation}
	introduced in Eqs.~(\ref{Valpha},\ref{VQzz}) to avoid a singularity at ${ r \to 0 }$.
	The parameter ${ {r}_{0} }$ in Eq.~(\ref{cutoff}) defines  an effective short range for the regularization.


The dipolar potential $V_{\rm dip}(r,\theta)$ is discontinuous at $r = s$. To remove this discontinuity, in Eq.~(\ref{V_dip}) we replace 
	\begin{eqnarray}
\frac{r_>}{r_<} \longrightarrow & \left\{\frac{r}{s} {f}_{a} (r) + \frac{s}{r} \left[ 1 - {f}_{a} (r) \right] \right\} \erf ( a r ) 	\label{eq_smooth_func}\\
	{r}_{>}  \longrightarrow &  s {f}_{a} (r) + r \left[ 1 - {f}_{a} (r) \right]
		\label{eq_smooth_r_sup}
	\end{eqnarray}
  with $f_{a} (r) = (1 + \exp[(r - s)/a])^{-1}$.

\section{Coupled-channel equations}
	\label{S:CC}
	
In the description of dipolar anions with the Hamiltonian (\ref{H_Born_Opp}), the coupled-channel formalism is well adapted to express the wave function of the system \cite{garrett70,garrett71,garrett80,garrett80a,garrett81,garrett82}. The eigenfunction of $H$ corresponding to
the total angular momentum ${ J }$ can be written as
\begin{equation}
		{ \Psi }^{J} = \sum_{c}{u}_{c}^{J} (r) { \Theta }_{ {\ell}_{c} {j}_{c}  }^{J},
		\label{eq_cc_wave_function}
	\end{equation}
where the index ${ c }$ labels the channel $(\ell,j)$, ${ {u}_{c}^{J} (r) }$ is the radial wave function of the valence electron,  ${ { \Theta }_{ {\ell}_{c} {j}_{c}  }^{J} }$ is the channel function, and  ${ \gras{j} + \gras{ \ell } = \gras{J} }$.
 Since the Hamiltonian is rotationally invariant, its eigenvalues are independent of the magnetic quantum
number ${ {M}_{J} }$, which will be omitted in the following.
	
In order to write the Schr\"odinger equation as a set of coupled-channel equations, the potential ${ V ( r , \theta ) }$ in Eqs.~(\ref{eq_main_pot} - \ref{VSR}) is expanded in multipoles:
	\begin{equation}
		V ( r , \theta ) = \sum_{ \lambda } {V}_{ \lambda} (r) {P}_{ \lambda } ( \cos\theta ),
	\end{equation}
where ${ {V}_{ \lambda} (r) }$ is the radial form factor and
	\begin{equation}
		{P}_{ \lambda } ( \cos\theta ) = \frac{ 4 \pi }{ 2 \lambda + 1 } {Y}_{ \lambda }^{ (mol) }( \hat{ \gras{s} } ) \cdot {Y}_{ \lambda }^{ (e) } ( \hat{ \gras{r} } ).
		\label{Plambda_sph_harmonics}
	\end{equation}

The matrix elements $\braket{ { \Theta }_{{\ell}_{c'}  {j}_{c'} }^{J} | {P}_{ \lambda } ( \cos\theta ) | {\Theta}_{ {\ell}_{c} {j}_{c}  }^{J} } $ are obtained by means of the standard angular momentum algebra \cite{fossez13}.
The resulting coupled-channel equations for the radial wave functions ${u}_{c}^{J} (r)$  can be written as:	
	\begin{eqnarray}
		&&\left[ \frac{ {d}^{2} }{ d{r}^{2} } - \frac{ { \ell }_{c} ( { \ell }_{c} + 1 ) }{ {r}^{2} } - \frac{ {j}_{c} ( {j}_{c} + 1 ) }{I} + {E}_{J} \right] {u}_{c}^{J} (r) \nonumber \\
		&=& \sum_{ c' } {V}_{ c c' }^{J} (r) {u}_{ c' }^{J} (r),
		\label{CC3}
	\end{eqnarray}
where ${ {E}_{J} }$ is the energy of the system and
	\begin{equation}
		{V}_{ c c' }^{J} (r) = \sum_{ \lambda } \braket{ { \Theta }_{ { \ell }_{ c' } {j}_{ c' }  }^{J} | {P}_{ \lambda } ( \cos\theta ) | { \Theta }_{{ \ell }_{c} {j}_{c}  }^{J} } {V}_{ \lambda } (r)
		\label{Vccp}
	\end{equation}
is the coupling potential.
	
\section{Berggren expansion method}
	\label{Berggren_basis_diag}

The Berggren expansion method for studies of© the bound states of dipolar anions has been introduced in Ref. \cite{fossez13}. In this method, the Hamiltonian is diagonalized  in a complete basis of single-particle (s.p.) states, the so-called Berggren ensemble \cite{berggren1968,Beggren1993,lind1993} which is generated by a finite-depth spherical one-body potential. The Berggren ensemble contains bound (${ b }$), decaying (${ d }$), and scattering (${ s }$) single-particle  states along the contour ${\cal L}^+_{\ell,j}$  for each considered partial wave $(\ell,j)$. For that reason, the Berggren ensemble is ideally suited to deal with weakly-bound and unbound  structures having large spatial extensions, such as halos, Rydberg states, or decaying resonances. For more details and recent applications of BEM in the many-body context, see Ref.~\cite{Mic09} and references cited therein.

While the finite-depth potential generating the Berggren ensemble can be chosen arbitrarily, to improve the convergence we take the diagonal part of the channel coupling potential ${ {V}_{cc'} (r) }$. The basis states ${ { \Phi }_{ k , c } (r) }$ are eigenstates of the spherical potential ${ {V}_{cc} (r) }$, which
are regular at origin and meet outgoing ($b,d$) and scattering ($s$) boundary conditions. Note that the wave number  ${ k }$ characterizing eigenstates
${ { \Phi }_{ k , c } (r) }$ is in general complex. The normalization of the bound states is standard, while that for the decaying states involves the exterior complex scaling \cite{Gyarmati1971,fossez13,Mic09}. The scattering states  are normalized to the Dirac delta function.

To determine Berggren ensemble, one calculates first the s.p. bound and resonance states of the generating s.p. potential for all chosen partial waves ${ ( \ell , j ) }$. Then, for each  channel ${ ( \ell , j ) }$, one selects the contour ${ { \cal L }^{+}_{ \ell , j } }$ in a fourth quadrant of the complex ${ k }$-plane. All ${ ( \ell , j ) }$-scattering states in this ensemble belong to ${ { \cal L }^{+}_{ \ell , j } }$. The precise form of ${ { \cal L }^{+}_{ \ell , j } }$ is unimportant providing that all selected s.p. resonances for a given ${ ( \ell , j ) }$ lie between this contour and the real ${ k }$-axis for ${ { \cal R } (k) > 0 }$. For each channel, the set of all resonant states and scattering states on ${ { \cal L }^{+}_{ { \ell }_{c} , {j}_{c} } }$ forms a complete s.p. basis. 
		
In the present study, each contour ${ { \cal L }^{+}_{ \ell , j } }$ is composed of three segments: the first one from the origin to ${ {k}_{ \text{peak} } }$ in the fourth quadrant of the complex ${ k }$-plane, the second one from ${ {k}_{ \text{peak} } }$ to ${ {k}_{ \text{middle} } }$ on the real ${ k }$-axis (${ { \cal R} (k) > 0 }$), and the third one from ${ {k}_{ \text{middle} } }$ to ${ {k}_{ \text{max} } }$ also on the real ${ k }$-axis. In all practical applications of the BEM, each contour ${ { \cal L }^{+}_{ \ell j } }$ is discretized and the Gauss-Legendre quadrature is applied. The  cutoff momentum ${ k = {k}_{max} }$ should be   sufficiently large to guarantee the completeness to a desired precision. The discretized  scattering states $\ket{ { \Phi }_{ n , c } }$ are renormalized  using the Gauss-Legendre weights. In this way,  the Dirac delta normalization of the scattering states is replaced by the usual Kronecker delta normalization. In this way, 
all ${ \ket{ { \Phi }_{ i , c } } }$ states can be treated on the same footing in
the  discretized Berggren completeness relation:
		\begin{equation}
			\sum_{ i = 1 }^{N} \ket{ { \Phi }_{ i , c } } \bra{ { \Phi }_{ i , c } } \simeq 1
			\label{discretized_Berggren_completeness},
		\end{equation}
where the ${ N }$ basis states include bound, resonance, and discretized scattering states for each considered channel ${ c }$. Finally, 
since the ${ {V}_{ c c' } (r) }$ decreases at least as fast as ${ {r}^{-2} }$, 
all the off-diagonal matrix elements of the  coupling potential can be computed by the means of
the complex scaling.

\section{Parameters of the BEM calculation}\label{parameters}

The parameters of the pseudo-potential for the HCN${ {}^{-} }$ anion are taken from Ref. \cite{garrett10}. These are:
	\begin{align}
		& { \alpha }_{0} = 15.27\,{a}_{0}^{3}, \nonumber\\ \nonumber
		& { \alpha }_{2} = 1.08 \,{a}_{0}^{3}, \\ \nonumber
		& {Q}_{ zz } = 3.28 \, e {a}_{0}^{2}, \\ \nonumber
		& I = 7.42 \times {10}^{4} \,{m}_{e} {a}_{0}^{2}, \\ \nonumber
		& {r}_{0} = 4.4 \,{a}_{0}, \\ \nonumber
		& {r}_{c} = 3.071622666 \,{a}_{0}, \\  \nonumber
		& {V}_{0} = 4.0 \, \text{Ry}, \\ \nonumber
		& s = 2.04 \,{a}_{0}, 
		\label{eq_pseudo_pot_param}
	\end{align}
and $a 	=a_0$.  The value of  ${ {r}_{c} }$ has been  adjusted to reproduce the experimental ground state  $({ {J}^{ \pi } = {0}^{+} })$ 
energy \cite{ard09}: ${E}^{ \text{exp} } ( {0}_{1}^{+} ) = - 1.1465789 \times {10}^{-4}$\,Ry.
For the dipolar moment of the molecule, we take the experimental value $\mu = 1.174$\,$e {a}_{0}$. In the following, we express $r$ in units of the Bohr radius $a_0$, $I$ in units of ${m}_{e} {a}_{0}^{2}$, and energy in Ry. The ${ {J}^{ \pi } = {1}^{-} }$ band head energy is also known experimentally, 
$E^\text{exp}(1_1^-) = - 8.8198377 \times 10^{-5}$\,Ry,
but no  adjustment of the model parameters has been attempted to fit the experimental value.

To achieve  stability of bound-state energies, the BEM calculations were carried out  by including all partial waves with $\ell \leq {\ell}_\text{max} = 9 $ and taking the optimized number of points ($N_{\rm C}=165$) on the complex contour with ${ {k}_{ \text{max} } = 6 \,{a}_{0}^{-1} }$ for each ${ {J}^{ \pi } }$. For all $(\ell, j)$ channels and all ${ {J}^{ \pi } }$-values, the complex contour ${ { \cal L }^{+}_{\ell, j} }$ is taken close to the real axis (${k}_{ \text{peak} } = 0.15 - i {10}^{-7}$, ${k}_{ \text{middle} } = 1.0$, and ${k}_{ \text{max} } = 6.0$; all in $a_0^{-1}$). Its precise form has been adjusted by looking at the convergence of bound state energies when changing the imaginary part of ${ {k}_{ \text{peak} } }$. Each segment of any contour ${ { \cal L }^{+}_{\ell, j} }$ is discretized with the same number of points ($N_{\rm C}/3=55$).

\section{Identification of the resonances}\label{s:reson}
	
The diagonalization of a complex-symmetric Hamiltonian matrix in BEM yields a set of eigenenergies which are the physical states (poles of the resolvent of the Hamiltonian) and a large number of complex-energy scattering states. The resonances are thus  embedded in a discretized continuum of scattering states and their identification is not trivial \cite{Mic02,Mic03}.
		
The eigenstates associated with resonances   should be stable with respect to changes of the contour \cite{Mic02,Mic03}. Moreover, 
their dominant channel wave functions should exhaust a large fraction of the real part of the norm. The norm of an eigenstate of the Hamiltonian is given by:
		\begin{equation}
			\sum_{c} \sum_{i} \braket{ { \Phi }_{ k , c } | {u}_{c} }^{2} =\sum_{c} {n}_{c} = 1,
			\label{eq_sum_channel_norms}
		\end{equation}
where  ${ {n}_{c} }$ the norm of the channel wave function. 		
In general, the norms of individual channel wave functions for resonances are complex numbers and their real parts are not necessarily positive definite. It may happen that if a large number of weak channels ${ \{ {c}_{i} \} }$ with small negative norms ${\cal R}({n}_{c_i}) < 0$ contribute to the resonance wave function, then the dominant channel ${ c }$ can have a norm ${ {n}_{c} > 1 }$. This does not come as a surprise as the channel wave functions have no obvious probabilistic interpretation.

To check the stability of BEM eigenstates, we varied the imaginary part of ${ {k}_{ \text{peak} } }$ from 0 to $-0.0001 {a}_{0}^{-1}$ in all partial-wave contours. Resulting contour variations change both real ${ \Delta \Re (E) } \ll \Re (E) $ and imaginary ${ \Delta \Im (E) }$ parts of the eigenenergies.
		\begin{table}[htb]
			\caption{Relative variation of the real part $\delta \Re (E) = \Delta \Re (E) / \Re (E)$ (in percent)  and imaginary part
			$\delta \Im (E) = \Delta \Im (E) / \Im (E)$ (in percent) of energies of twenty  ${ {J}^{ \pi } = {2}^{+} }$ resonances  with the change of  
			${k}_\text{peak}$. All energies are in Ry. The numbers in parentheses denote powers of 10.}
			\begin{ruledtabular}
				\begin{tabular}{ccccc}
					resonance & ${ \Re (E) }$ & $\delta \Re (E)$ & ${ \Im (E) }$  & $\delta \Im (E)$ \\
					\hline \\[-6pt]
					1  & 2.51(-5) & 2.47(-1) & -9.68(-6)  & 2.09(-1) \\
					2  & 2.69(-4) & 1.29(-4) & -3.45(-10) & 1.32(+1) \\
					3  & 2.77(-4) & 1.37(-5) & -3.58(-9)  & 1.56(+1) \\
					4  & 3.55(-4) & 5.61(-4) & -7.20(-7)  & 1.60     \\
					5  & 3.67(-4) & 3.70(-4) & -1.21(-6)  & 1.78     \\
					6  & 3.96(-4) & 3.52(-3) & -2.34(-6)  & 4.55(-1) \\
					7  & 3.98(-4) & 2.07(-2) & -5.05(-5)  & 6.19(-2) \\
					8  & 4.25(-4) & 6.02(-3) & -1.04(-4)  & 3.02(-2) \\
					9 & 6.48(-4) & 9.70(-5) & -6.72(-7)  & 1.42     \\
					10 & 6.60(-4) & 6.86(-4) & -8.32(-7)  & 2.52     \\
					11 & 6.81(-4) & 6.77(-3) & -1.19(-5)  & 7.41(-1) \\
					12 & 6.86(-4) & 9.86(-4) & -1.60(-6)  & 1.55     \\
					13 & 7.40(-4) & 5.05(-3) & -6.68(-5)  & 3.85(-2) \\
					14 & 9.80(-4) & 7.89(-4) & -7.86(-7)  & 1.45(+1) \\
					15 & 1.05(-3) & 4.80(-5) & -6.22(-7)  & 1.39     \\
					16 & 1.06(-3) & 1.87(-4) & -8.54(-7)  & 2.66     \\
					17 & 1.07(-3) & 1.82(-3) & -5.60(-6)  & 1.10     \\
					18 & 1.09(-3) & 4.00(-4) & -4.89(-7)  & 7.67     \\
					19 & 1.11(-3) & 8.05(-4) & -1.66(-6)  & 9.61     \\
					20 & 1.14(-3) & 2.28(-3) & -2.71(-5)  & 1.31(-1) \\
				\end{tabular}
			\end{ruledtabular}
			\label{tab_var_real_im_E}
		\end{table}
The precision of the resonance-identification method is  assessed by looking at the ratio ${ \Delta \Im (E) / \Im (E) }$, which is in the range ${ [0.001 , 0.3] }$ for the resonance states. As an example, the eigenvalues of ${ {J}^{ \pi } = {2}^{+} }$ resonant states are listed in Table \ref{tab_var_real_im_E}. 	
\begin{figure}[htb]
\includegraphics[width=0.8\columnwidth]{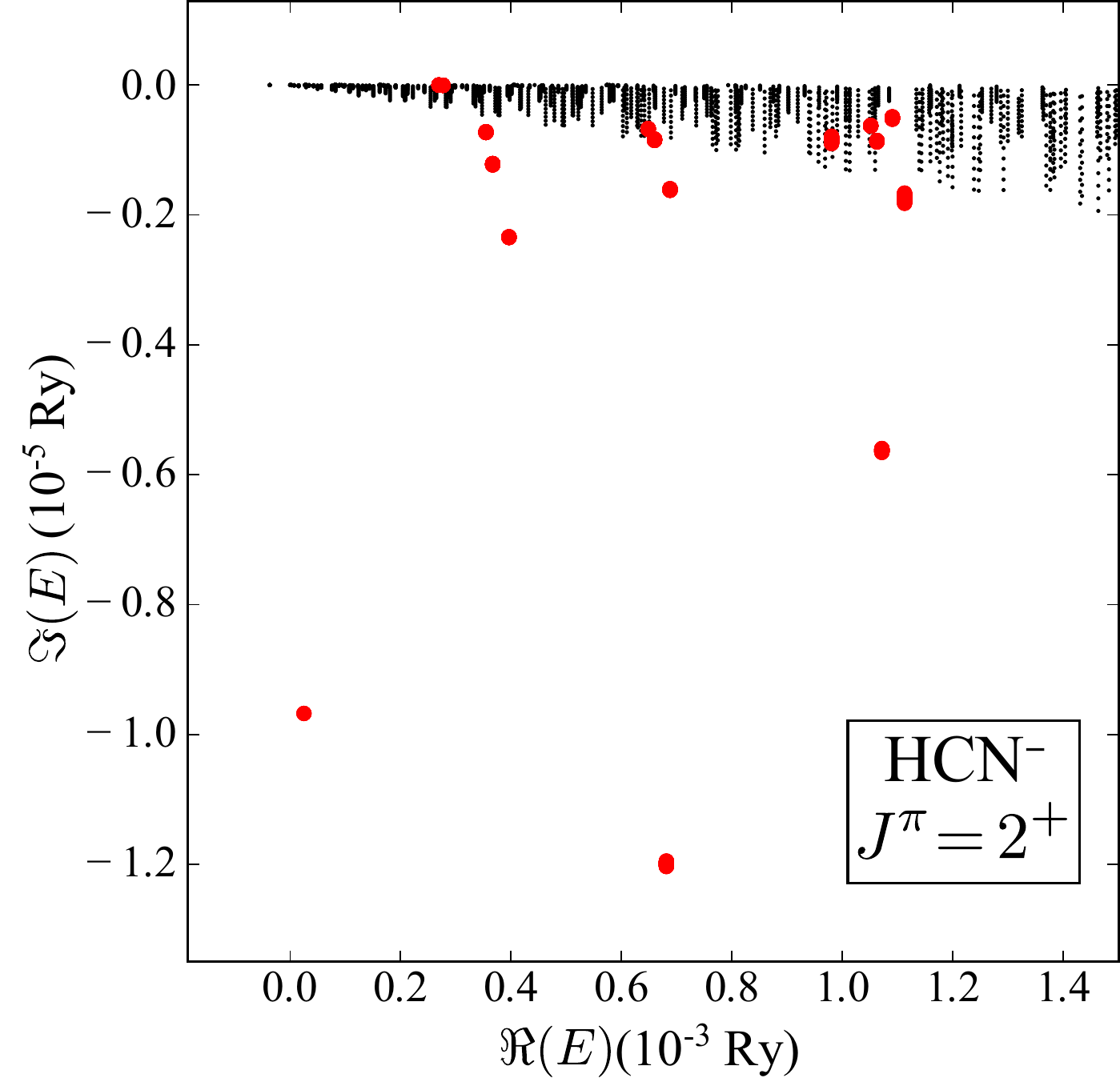}
\caption{(Color online) Illustration of the stability of the energies of the ${ {J}^{ \pi } = {2}^{+} }$ resonant states of HCN$^-$  listed in Table~\ref{tab_var_real_im_E} (large dots) when the non-resonant scattering contour is shifted. Here, the imaginary part of ${ {k}_{ \text{peak} } }$ was varied  from 0 to $-0.0001 {a}_{0}^{-1}$. As a comparison, non-resonant eigenenergies are marked with tiny dots and exhibit significant shifts.} 
\label{fig_plot_acc_hcn_2_all}
		\end{figure}
\begin{figure}[htb]
			\includegraphics[width=0.8\columnwidth]{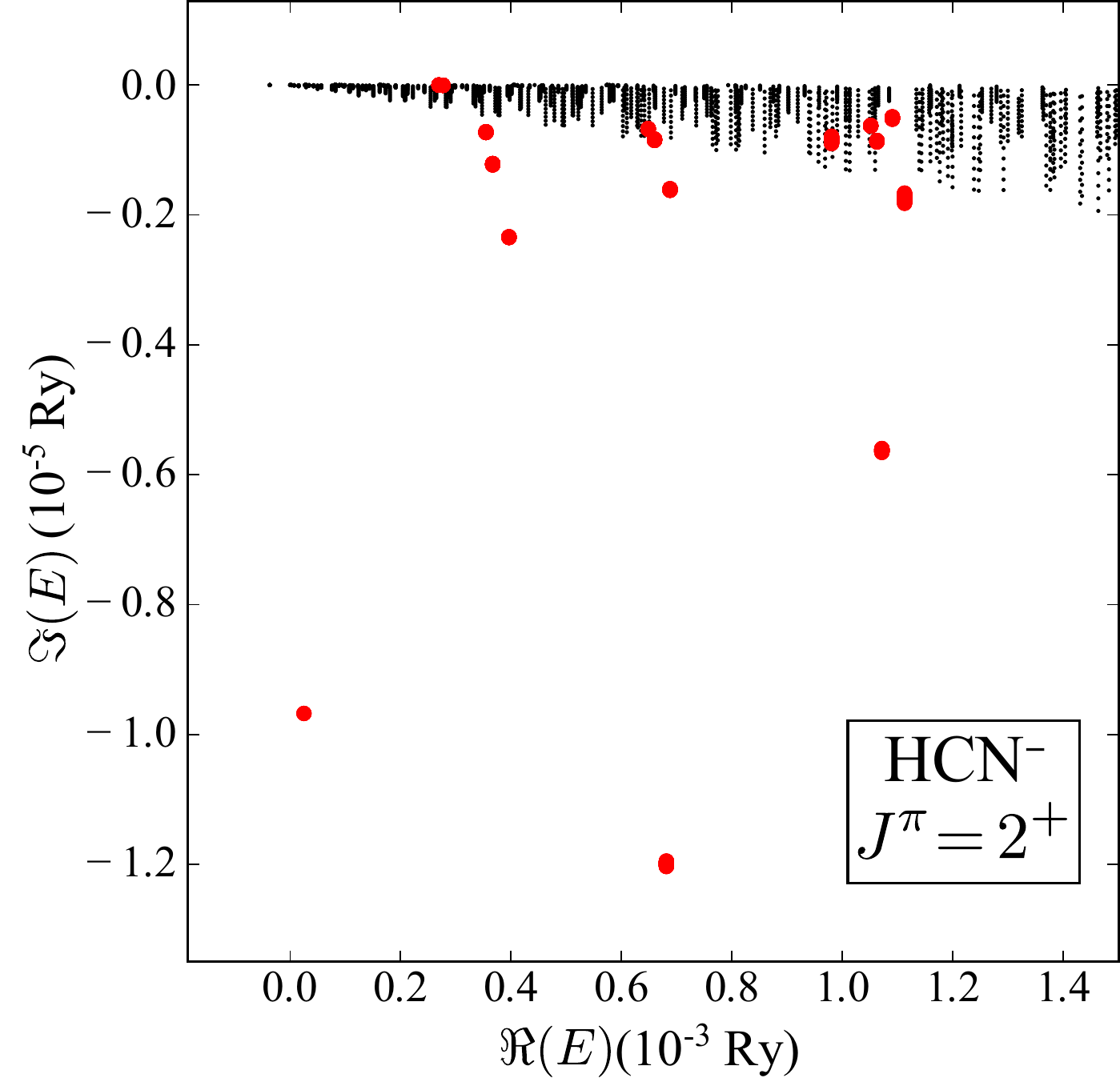}
			\caption{(Color online) Similar as in
			Fig.~\ref{fig_plot_acc_hcn_2_all} but zoomed in on the two threshold resonances (states 2 and 3 in Table~\ref{tab_var_real_im_E}). 
			Here, the real part of ${ {k}_{ \text{peak} } }$ is also varied  from 0.14\,$a_0^{-1}$ to 0.16\,$a_0^{-1}$.}
			\label{fig_plot_acc_hcn_2}
\end{figure}
It is seen that the relative variations of $\Re (E)$  are always smaller than 1\%, while the relative variations of $\Im(E)$  can reach ${\sim }$15\%. Moreover, values of ${ \Delta \Im (E) / \Im (E) }$ for different resonant states can differ by three orders of magnitude. In general, a better stability of the BEM eigenstates and, i.e., smaller values of ${ \Delta \Im (E) / \Im (E) }$, is found for those eigenstates, which have several channel wave functions contributing significantly to the total norm. A typical accumulation of eigenenergies when changing the contour is shown in Fig.~\ref{fig_plot_acc_hcn_2_all}. One can see that the non-resonant  states  do not exhibit the degree of stability that is typical of resonant states. It is interesting to notice that several resonant states are found fairly away from the region of  non-resonant eigenstates. The stability of resonant eigenstates persists if the real part of ${ {k}_{ \text{peak} } }$ is varied from 0.14\,$a_0^{-1}$ to 0.16\,$a_0^{-1}$. In this case, the relative variations of the real part of the eigenstate energies dominate as can be seen in Fig. \ref{fig_plot_acc_hcn_2} for the two near-threshold resonances labeled 2 and 3 in Table~\ref{tab_var_real_im_E}.

\begin{figure}[htb]
\includegraphics[width=0.8\columnwidth]{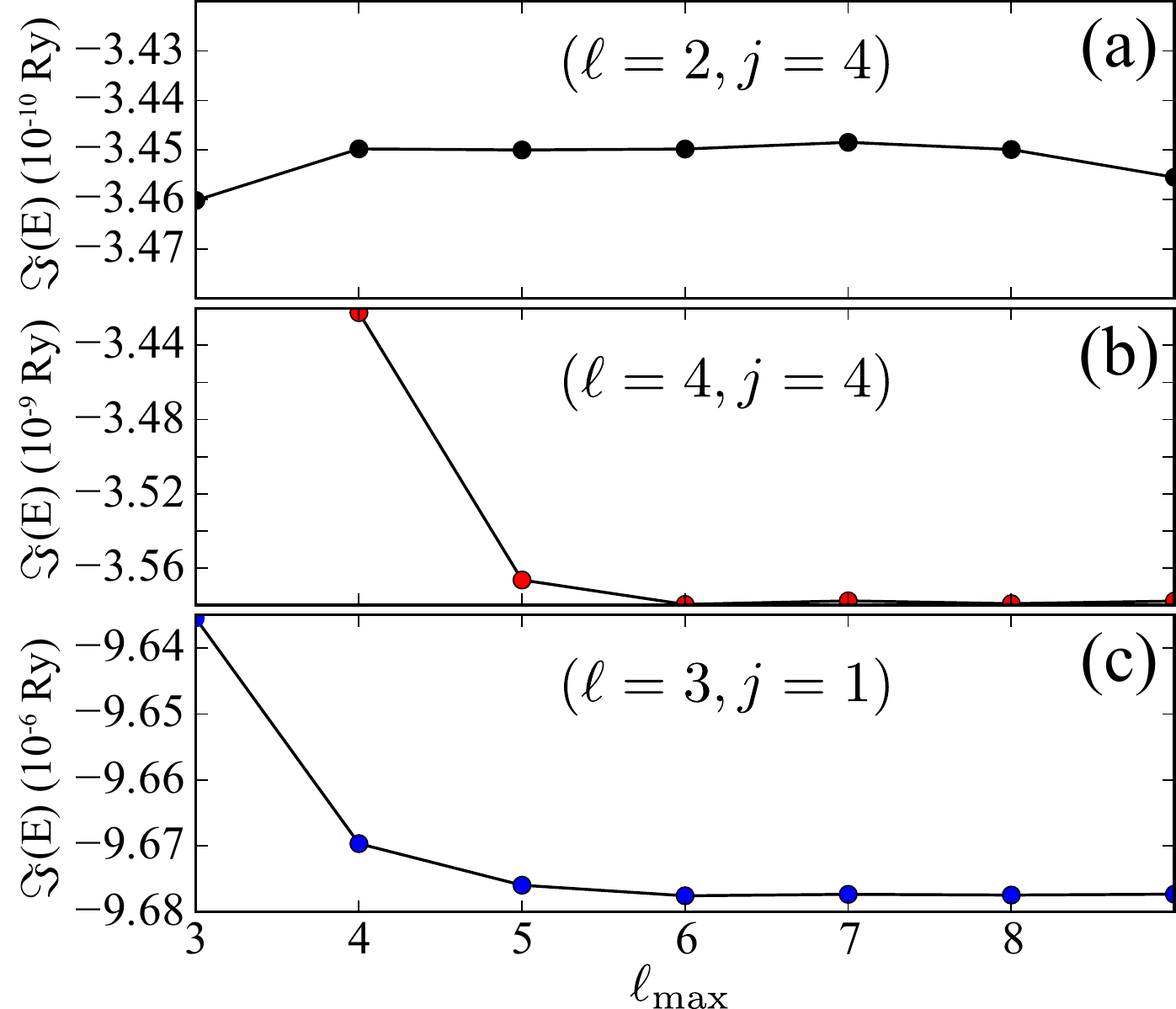}
			\caption{(Color online) The convergence of $\Im(E)$ for ${ {J}^{ \pi } = {2}^{+} }$ resonances 2 (a), 3 (b), and 1 (c) of 
			Table~\ref{tab_var_real_im_E} as a function of ${ { \ell }_{ \text{max} } }$. The quantum numbers $(\ell, j)$ of the dominant channel are indicated.}
			\label{fig_plot_var_lmax_hcn_2_three_res}
		\end{figure}

In order to demonstrate that the identified resonances are stable with respect 
to  $\ell_\text{max}$, in Fig.~\ref{fig_plot_var_lmax_hcn_2_three_res} we show the energy convergence for states 1-3 of Table~\ref{tab_var_real_im_E}. In general, $\Im(E)$  is significantly more sensitive than $\Re(E)$ with respect to the addition of channels with higher $\ell$- and $j$-values. It is seen that $\Im(E)$ for resonances with the dominant channels $( \ell = 4 , j = 4 )$  and $( \ell = 3 , j = 1 )$ are converged already for $\ell_{\rm max} \geq 6$.
The convergence for the narrow resonance with the dominant channel $(\ell = 2 , j = 4 )$ shown in Fig.~\ref{fig_plot_var_lmax_hcn_2_three_res}(a) is also excellent, considering that in this case $\Im(E)$ is of the order of $10^{-10}$\,Ry, which is close to the limit of a numerical precision of our BEM calculations.

\section{Intrinsic density}\label{s:density}

It is instructive to present the density of the valence electron in the body-fixed frame. This can easily be done in the strong coupling scheme 
of the  particle-plus-rotor model \cite{Vleck51,Her66,bohr98a}, which is usually formulated in the $K$-representation associated with the intrinsic frame. 
Here, $K_J=K_\ell +K_j$ is the projection of the total angular momentum on the symmetry axis of the molecule. Of particular interest
is the  adiabatic limit of ${ I \to \infty }$, where all ${ {J}^{ \pi } }$ members of a rotational band collapse at the band-head, i.e.,  they  all can be associated with one intrinsic configuration. The $K$-representation is useful to visualize wave functions, group states with different $J$-values into rotational bands,  and interpret the results in terms of Coriolis mixing  \cite{kruppa99,kruppa03,Kru04,esbensen00,davids04,Chernov05}.

In the body-fixed frame, the density of the valence electron in the state $J^\pi$ is axially-symmetric and can be decomposed as:
\begin{equation}
	\rho_J ( r , \theta ) = \sum_{ K_J } \rho_{ J K_J } (r,\theta),
	\label{eq_int_den_fixed_orientation_0}
\end{equation}
where  $(r,\theta)$ stand for the polar coordinates of the electron in the intrinsic frame, and the $K_J$-components of the density are:
\begin{eqnarray}
	\rho_{ J K_J } (r,\theta) &=& \sum_{ \ell , \ell '}  \sum_{j}  \frac{2j+1 }{ 2 J + 1 } \braket{ \ell  K_J j 0 | J  K_J } \braket{ \ell'  K_J  j  0 | J  K_J }  \nonumber \\
&\times & \frac{ u^J_{ \ell j } (r) ^* }{r} \frac{ {u}^J_{ \ell ' j } (r) }{r}  Y_\ell^{K_J*}(\theta,0)\, Y_{\ell '}^{K_J}(\theta,0).
	\label{eq_int_density_KJ_proj}
\end{eqnarray}

If all $K_J$-components except one vanish in Eq.~(\ref{eq_int_den_fixed_orientation_0}), the adiabatic strong-coupling limit is reached and $K_J$ becomes a good quantum number. In this particular case, $\rho_{ J{K}_{J} }$ can be identified as the intrinsic electronic density in the dipole-fixed reference frame. 
To quantify the degree of $K_J$-mixing, it is convenient to introduce the normalization amplitudes:
\begin{equation}
n_{J K_J} = \sum_{\ell,j} \frac{2j+1}{2J+1} \braket{\ell K_J j 0|J K_J}^{2} \int | u_{\ell j}^J (r) |^2\,dr.
	\label{nJK}
\end{equation}
Due to (\ref{eq_sum_channel_norms}), $n_{J K_J}$   fullfil the normalization condition:
\begin{equation}
\sum_{K_J} n_{J K_J} = 1.
	\label{nJKnorm}
\end{equation}

\section{Results of BEM calculations}\label{results}

Predicted energy spectra of HCN${ {}^{-} }$ with ${J}^{ \pi } = {0}^{+}, {1}^{-}, {2}^{+}, {3}^{-}, {4}^{+}$ and ${5}^{-}$  are shown in 
Table~\ref{tab_spectra}. One may notice that the calculated energy of the ${ {1}^{-} }$ band head, $E ( {1}_{1}^{-} ) = -8.96 \times {10}^{-5}$\,Ry, is close to the experimental value $E^\text{exp} (1_1^-) = - 8.82 \times {10}^{-5}$\,Ry. Moreover, consistently with earlier Refs. \cite{garrett10,ard09}, we do not find a ${ {J}^{ \pi } = {3}^{-} }$ bound state.

\begin{table*}[htb]
\caption{\label{tab_spectra} Predicted complex energies (in Ry) of bound and resonance  ${ {0}^{+}, {1}^{-}, {2}^{+}, {3}^{-} }$, ${ {4}^{+} }$, and ${ {5}^{-} }$ states of the HCN${ {}^{-} }$ dipolar anion. Numbers in the parentheses denote powers of 10.}
\begin{ruledtabular}
	\begin{tabular}{|c|c|c|c|c|c|c|}
		\multicolumn{1}{|c}{state} & \multicolumn{1}{c}{${ E ( {0}^{+} ) }$} & \multicolumn{1}{c}{${ E ( {1}^{-} ) }$}  & \multicolumn{1}{c}{${ E ( {2}^{+} ) }$} & \multicolumn{1}{c}{${ E ( {3}^{-} ) }$} & \multicolumn{1}{c}{${ E ( {4}^{+} ) }$} & \multicolumn{1}{c|}{${ E ( {5}^{-} ) }$} \\
		\hline \\[-6pt]
		1  & -1.15(-4)	 		& -8.96(-5) 			& -3.69(-5) 			& 3.89(-8) -i 1.06(-8) & 2.70(-5) -i 5.55(-9) & 8.09(-5) -i 3.08(-9) \\
		2  & 7.62(-5) -i 3.79(-6) 	& 2.70(-5) -i 9.98(-10) 	& 2.51(-5) -i 9.68(-6) 		& 2.63(-4) -i 1.88(-6) & 1.84(-4) -i 2.02(-6) & 1.33(-4) -i 2.02(-6) \\
		3  & 9.35(-4) -i 9.69(-5) 	& 8.12(-5) -i 7.04(-7) 		& 2.69(-4) -i 3.45(-10) 	& 3.03(-4) -i 9.25(-6) & 2.25(-4) -i 2.47(-5) & 1.63(-4) -i 3.71(-5) \\
		4  & 1.09(-3) -i 1.24(-5) 	& 1.62(-4) -i 4.77(-10) 	& 2.77(-4) -i 3.58(-9) 		& 4.99(-4) -i 1.28(-6) & 3.65(-4) -i 1.40(-6) & 2.56(-4) -i 1.87(-6) \\
		5  & 1.11(-3) -i 4.06(-4) 	& 4.88(-4) -i 7.04(-7) 		& 3.55(-4) -i 7.20(-7) 		& 5.32(-4) -i 1.01(-6) & 3.99(-4) -i 1.43(-6) & 2.91(-4) -i 1.85(-6) \\
		6  & 1.14(-3) -i 1.62(-5) 	& 5.00(-4) -i 1.02(-6) 		& 3.67(-4) -i 1.21(-6) 		& 5.69(-4) -i 1.25(-4) & 4.23(-4) -i 1.26(-4) & 3.03(-4) -i 1.22(-4) \\
		7  & 1.16(-3) -i 2.19(-4) 	& 5.28(-4) -i 1.65(-6) 		& 3.96(-4) -i 2.34(-6) 		& 8.20(-4) -i 1.17(-5) & 6.58(-4) -i 9.78(-7) & 4.94(-4) -i 1.03(-6) \\
		8  & 1.19(-3) -i 1.96(-5) 	& 5.34(-4) -i 3.13(-5) 		& 3.98(-4) -i 5.05(-5) 		& 8.80(-4) -i 2.96(-7) & 6.91(-4) -i 3.44(-7) & 5.28(-4) -i 3.62(-7) \\
		9  & 1.27(-3) -i 2.13(-5) 	& 5.71(-4) -i 9.11(-5) 		& 4.25(-4) -i 1.04(-4) 		& 9.39(-4) -i 9.91(-5) & 6.92(-4) -i 1.07(-5) & 5.67(-4) -i 9.80(-5) \\
		10 & 1.31(-3) -i 3.45(-4) 	& 6.71(-4) -i 3.31(-4) 		& 6.48(-4) -i 6.72(-6) 		& 1.07(-3) -i 3.55(-4) & 7.40(-4) -i 1.01(-4) & 5.92(-4) -i 9.87(-6) \\
		11 & 1.43(-3) -i 5.64(-6) 	& 8.37(-4) -i 6.53(-7) 		& 6.60(-4) -i 8.32(-7) 		& 1.16(-3) -i 1.24(-5) & 8.66(-4) -i 3.38(-4) & 6.82(-4) -i 3.14(-4) \\
		12 & 1.84(-3) -i 1.10(-5) 	& 8.48(-4) -i 8.03(-7) 		& 6.81(-4) -i 1.19(-5) 		& 1.30(-3) -i 7.87(-7) & 9.75(-4) -i 1.15(-5) & 8.21(-4) -i 1.23(-5) \\
		13 & 3.35(-3) -i 1.42(-4) 	& 8.63(-4) -i 8.45(-6) 		& 6.88(-4) -i 1.60(-6) 		& 1.34(-3) -i 1.09(-7) & 1.06(-3) -i 7.83(-7) & 8.44(-4) -i 7.67(-7) \\
		14 & 3.68(-3) -i 3.26(-5) 	& 8.76(-4) -i 9.82(-7) 		& 7.40(-4) -i 6.68(-5) 		& 1.41(-3) -i 7.12(-5) & 1.09(-3) -i 1.16(-7) & 8.78(-4) -i 1.14(-7) \\
		15 & 4.23(-3) -i 3.47(-4) 	& 9.34(-4) -i 5.08(-5) 		& 9.80(-4) -i 7.86(-7) 		& 1.56(-3) -i 3.54(-4) & 1.16(-3) -i 7.50(-5) & 9.34(-4) -i 7.38(-5) \\
		16 & 4.60(-3) -i 4.45(-5) 	& 1.05(-3) -i 3.13(-4) 		& 1.05(-3) -i 6.22(-7) 		& 1.61(-3) -i 1.41(-5) & 1.30(-3) -i 3.37(-4) & 1.06(-3) -i 3.13(-4) \\
		17 &  				& 1.17(-3) -i 7.06(-7)  	& 1.06(-3) -i 8.54(-7) 		& 1.65(-3) -i 7.83(-4) & 1.37(-3) -i 1.24(-5) & 1.16(-3) -i 1.08(-5) \\
		18 &  				& 1.30(-3) -i 3.00(-4)  	& 1.07(-3) -i 5.60(-6) 		& 2.17(-3) -i 1.60(-5) & 1.67(-3) -i 4.88(-5) & 1.30(-3) -i 6.64(-7) \\
		19 &  				& 1.30(-3) -i 1.41(-6)  	& 1.09(-3) -i 4.89(-7) 		& 2.24(-3) -i 7.85(-4) & 1.84(-3) -i 3.41(-4) & 1.40(-3) -i 4.89(-5) \\
		20 &  				& 1.62(-3) -i 5.82(-7)  	& 1.11(-3) -i 1.66(-6) 		&  		       & 1.88(-3) -i 1.44(-5) & 1.55(-3) -i 3.19(-4) \\
		21 &  				& 1.78(-3) -i 2.83(-4)  	& 1.14(-3) -i 2.71(-5) 		&  		       & 1.94(-3) -i 7.63(-4) & 1.61(-3) -i 1.27(-5) \\
		22 &  				&  				&  				&  		       & 2.49(-3) -i 1.64(-5) & 1.66(-3) -i 7.36(-4) \\
		23 &  				&  				&  				&  		       & 2.58(-3) -i 7.73(-4) & 1.96(-3) -i 3.15(-5) \\
		24 &  				&  				&  				&  		       & 		      & 2.14(-3) -i 3.29(-4) \\
		25 &  				&  				&  				&  		       & 		      & 2.17(-3) -i 1.46(-5) \\
		26 &  				&  				&  				&  		       & 		      & 2.25(-3) -i 7.44(-4) \\
		27 &  				&  				&  				&  		       & 		      & 2.84(-3) -i 1.67(-5) \\
		28 &  				&  				&  				&  		       & 		      & 2.94(-3) -i 7.61(-4) \\
	\end{tabular}
\end{ruledtabular}
\end{table*}
%
\begin{figure}[htb]
\includegraphics[width=0.8\columnwidth]{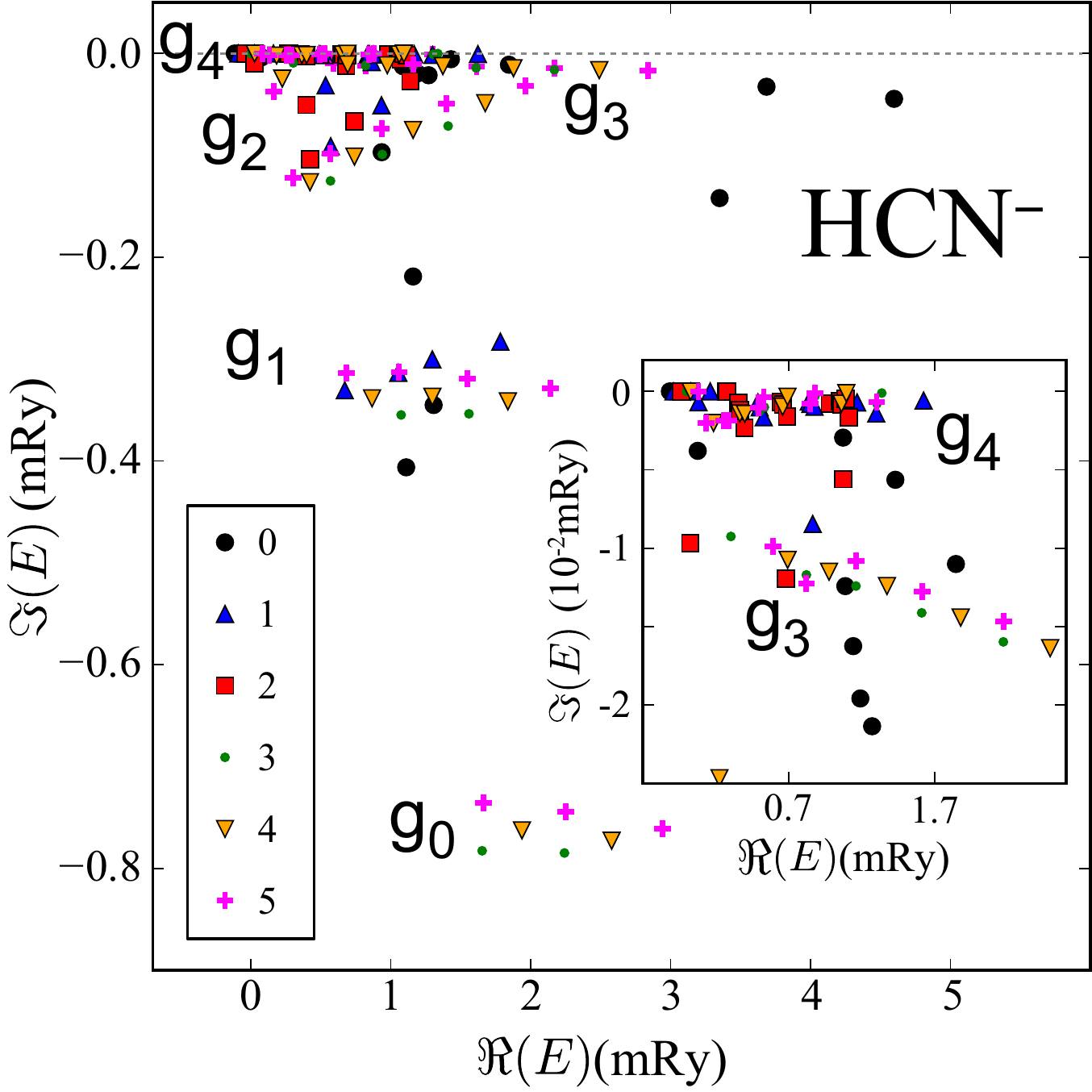}
\caption{(Color online) Predicted energies of the HCN$^-$ dipolar anion for ${ {J}^{ \pi } = {0}^{+} }$, ${ {1}^{-} }$, ${ {2}^{+} }$, ${ {3}^{-} }$, ${ {4}^{+} }$, and ${ {5}^{-} }$ states in the complex-energy plane. Based on their complex energies, these states can be organized into five groups labelled $g_0$ to $g_4$. Bound states and near-threshold resonances belonging to $g_4$  and narrow resonances of $g_3$ are shown in the insert.}
\label{fig_spectra}
\end{figure}
The states listed in Table \ref{tab_spectra} are plotted in 
Fig.~\ref{fig_spectra} in the  complex energy plane. These states can be assembled  according to their decay widths into  five groups labelled $g_0$-$g_4$.
The group 4 contains bound states and very narrow threshold resonances of the dipolar anion.
Narrow resonances are contained in groups 3 and 2 while broader states form groups 1 and 0. The characterization of the resonance spectra of
HCN$^-$ in terms of groups  $g_0$-$g_4$ will be provided below.

\subsection{Adiabatic limit}
	
\begin{figure}[htb]
	\includegraphics[width=0.8\columnwidth]{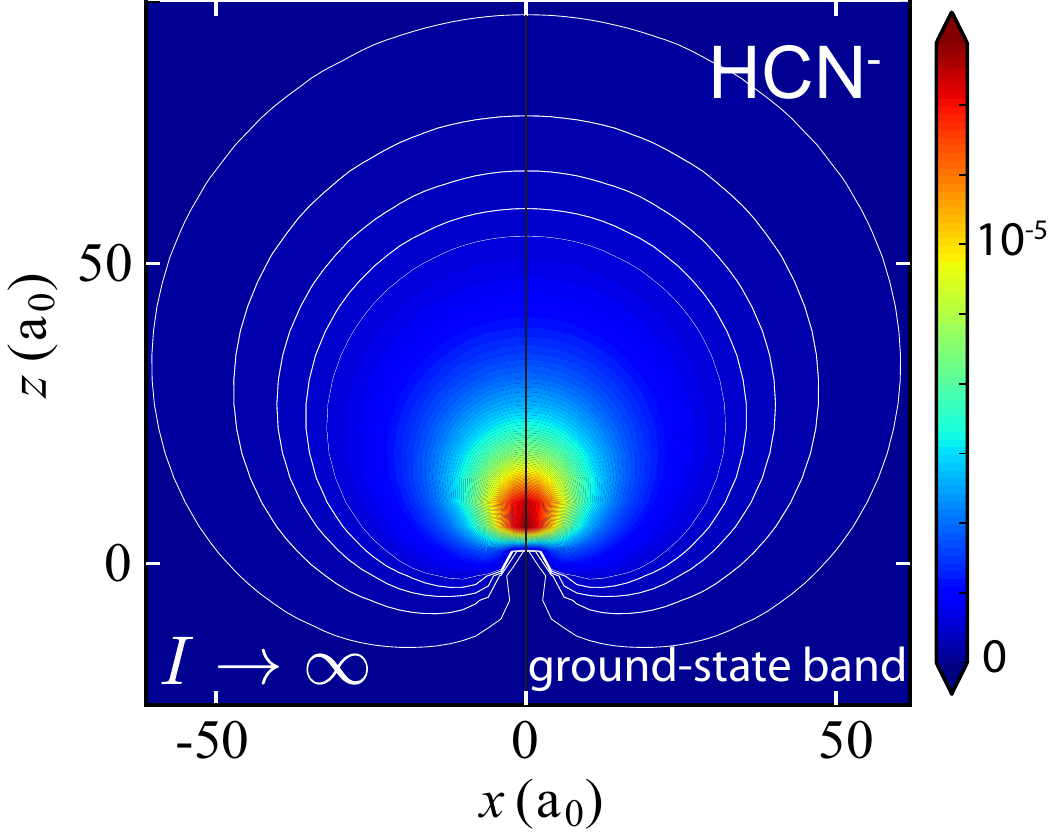}
\caption{(Color online) The intrinsic density of the valence electron in HCN$^-$ in  the ground-state rotational band $J^\pi= 0_1^+, 1_1^-, 2_1^+, 3_1^-, \dots $ (All densities are in $a_0^{-1}$.)}
\label{fig_plot_map_hcn_0_I_inf_K0}
\end{figure}
To check the numerical accuracy of the adiabatic approximation, we computed the energies of the lowest  states of  HCN${ {}^{-} }$ in the adiabatic limit 
of ${ I \to \infty }$ (in practice, $I=10^{16}$\,$ {m}_{e} {a}_{0}^{2}$). In this limit, which can be associated with the extreme strong coupling regime, 
$K_J$ becomes a good quantum number and energies of all band members $J=K_J, K_J+1, K_J+2,\dots$ collapse at the bandhead $E_{J=K_J}$. 
In our calculations, the maximum energy difference  between the members of the ground-state band (${J}^{ \pi } = 0_1^+, 1_1^-, 2_1^+, 3_1^-, 4_1^+, 5_1^-$)  
is $1.5 \times 10^{-7}$\,Ry, which is better than 0.1\% of the  energy of the ${ {0}^{+} }$ state ($E=-1.2308 \times 10^{-4}$\,Ry). 
We can conclude, therefore, that the members of the ground-state rotational band are practically degenerate in the adiabatic limit.

\begin{figure}[htb]
\includegraphics[width=0.8\columnwidth]{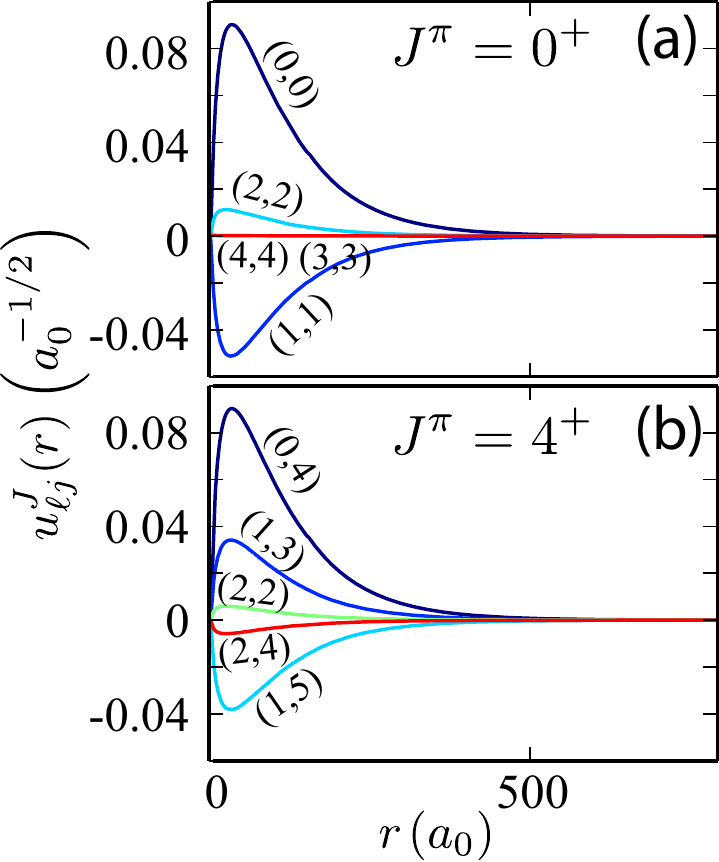}
\caption{(Color online) Channel wave functions $(\ell,j)$ of the $J^\pi = 0_1^+$ (a)  and $4_1^+$ (b) members of the ground-state rotational band in HCN$^-$ in the adiabatic limit.}
\label{fig_plot_wfs_hcn_0_I_inf}
\end{figure}

Figure~\ref{fig_plot_map_hcn_0_I_inf_K0} illustrates  the intrinsic density for the ground-state band in the adiabatic limit ($I \to \infty;~ K_{J} = 0$).  
The intrinsic densities for all band members are numerically identical even though the associated wave functions in the laboratory system are different, 
see Fig.~\ref{fig_plot_wfs_hcn_0_I_inf}.
The strongly asymmetric shape of electron's distribution reflects the attraction/repulsion between the electron and positive/negative charge of the dipole (for other illustrative examples, see Refs.~\cite{desfrancois96,Desfrancois04,Burrow06,Simons08,ard09}).

\begin{figure}[htb]
	\includegraphics[width=0.7\columnwidth]{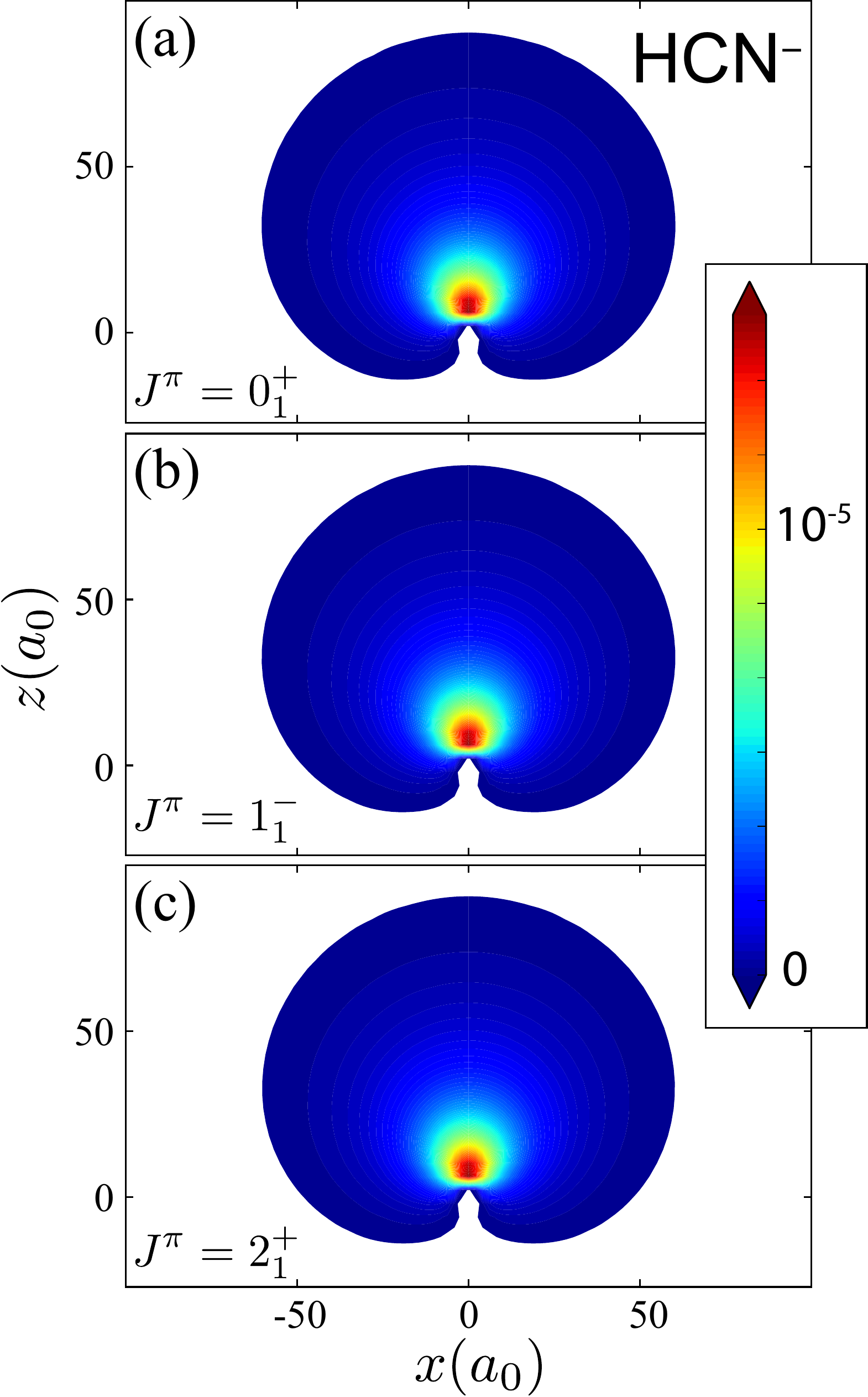}
	\caption{(Color online) Density (\ref{eq_int_den_fixed_orientation_0}) of the valence electron in the  bound states $ J^{\pi} = 0_1^+$ (a), $1_1^-$ (b), and $2_1^+$ (c) of  HCN$^-$. (All densities are in $a_0^{-1}$.) }
	\label{plot_map_hcn_0_0_K_0}
\end{figure}

We found that the  density representation given by Eq.~(\ref{eq_int_den_fixed_orientation_0}) can also  be useful in the non-adiabatic case, with finite 
moment of inertia, to assign members of rotational bands. This is illustrated in Fig.~\ref{plot_map_hcn_0_0_K_0} which shows
the density (\ref{eq_int_den_fixed_orientation_0}) for the  bound states $J^{\pi} = 0_1^+$, $1_1^-$, and $2_1^+$ of  HCN$^-$. Despite the fact that the strong coupling limit does not strictly apply in this case, distributions are practically identical and  close to the intrinsic density  displayed in 
Fig.~\ref{fig_plot_map_hcn_0_I_inf_K0}.

\subsection{Rotational bands}

\begin{figure}[htb]
\includegraphics[width=0.8\columnwidth]{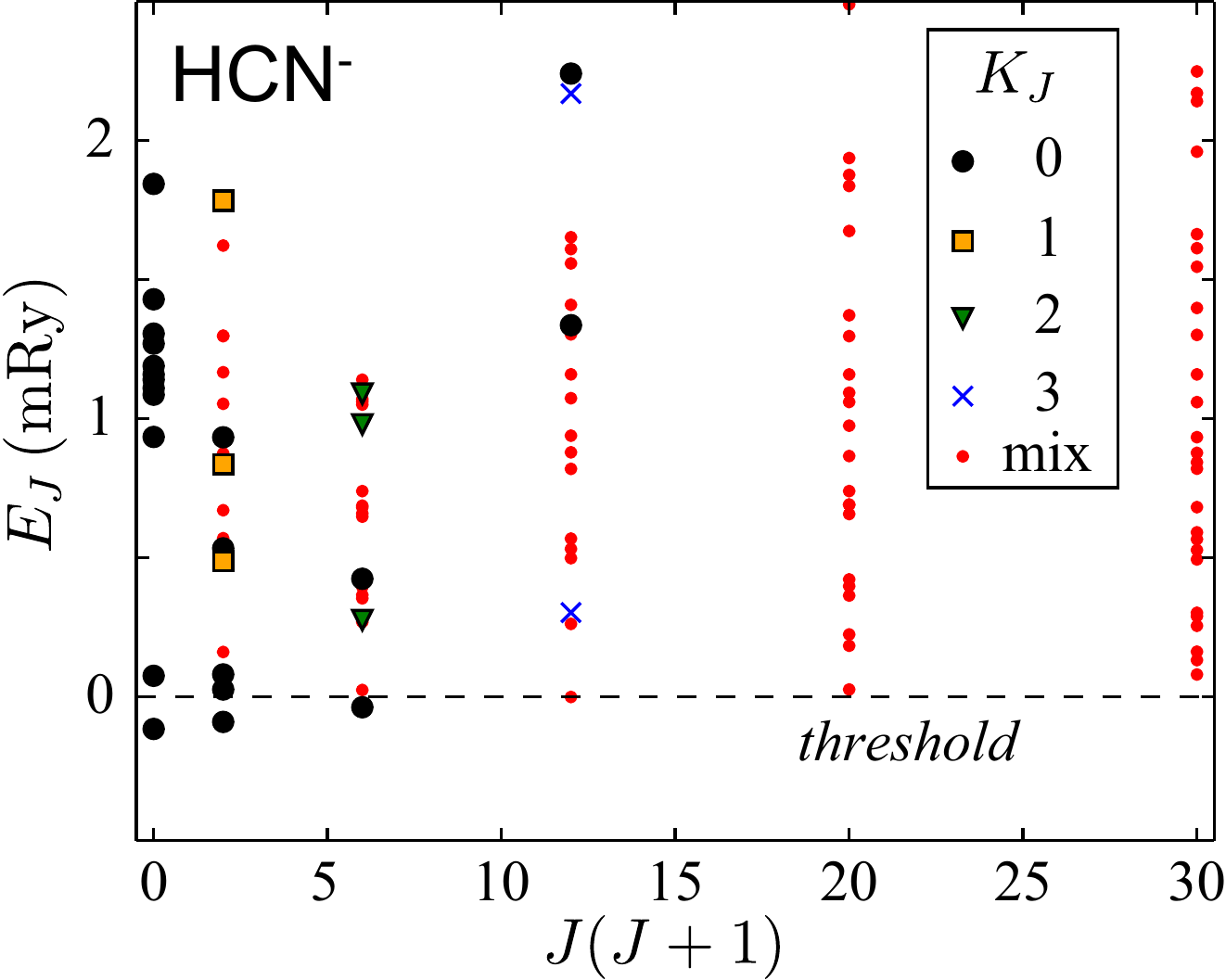}
\caption{(Color online) Energy  spectrum of the HCN$^-$ anion for ${ {J}^{ \pi } = {0}^{+} }$, ${ {1}^{-} }$, ${ {2}^{+} }$, ${ {3}^{-} }$, 
${ {4}^{+} }$, and ${ {5}^{-} }$  shown as a function of ${ J ( J + 1 ) }$. The dominant $K_J$-component (\ref{nJK}) is indicated. 
If several components are present, the state is marked as ``mix".}
\label{fig_plot_diagram_JJp1}
\end{figure}

Excitation energies of the lowest-energy resonant (i.e., bound and resonance)  states are plotted in Fig.~\ref{fig_plot_diagram_JJp1} as a function of ${ J ( J + 1 ) }$. The ${ {J}^{ \pi } = {0}^{+} }$, ${ {1}^{-} }$, ${ {2}^{+} }$ bound states form a ${ {K}_{J} = 0 }$ rotational band as evidenced by their intrinsic densities shown in Fig.~\ref{plot_map_hcn_0_0_K_0}.
Another ${ {K}_{J} = 0 }$ rotational band is built upon the ${ {0}_{2}^{+} }$ resonance. According to Table~\ref{tab_spectra}, a $1_2^-$ member 
of this band has a decay width that is  reduced by over three orders of magnitude as compared to that of the $0^+_2$ bandhead. We predict other very narrow resonances as well. Among them, the $2^+_4$ state has $K_J=2$ while $1_4^-$ and $2^+_3$ resonances have a mixed character.

\begin{figure}[htb]
	\includegraphics[width=0.7\columnwidth]{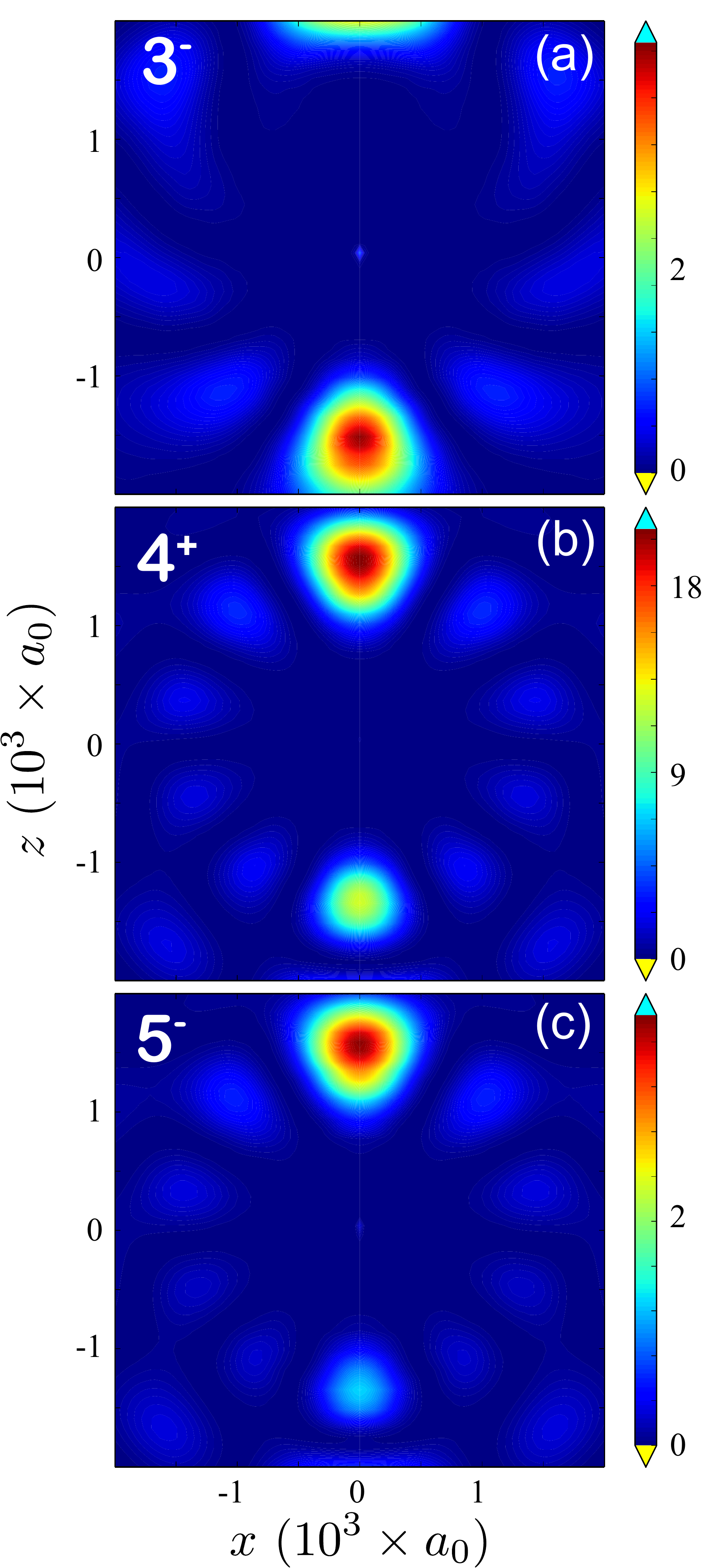}
	\caption{(Color online) Similar as in Fig.~\ref{plot_map_hcn_0_0_K_0} but for $\rho_{ J K_J=0 } (r,\theta)$ (in $10^{-15}\,a_0$)  in 
	(a) $3_1^-$, (b) $4_1^+$, and (c) $5^-_1$.}
	\label{fig_map_hcn_3_0__4_3__5_8_K_0}
\end{figure}

As can be judged by results displayed in   Fig.~\ref{fig_plot_diagram_JJp1}, except for few states with well defined $K_J$-values, majority of  resonances  
are strongly $K_J$-mixed. Consequently, an identification of other rotational bands in the continuum, based on the concept of intrinsic density, is not straightforward. This is true, in particular for the supposed higher-$J$ members of the ground-state band.
Figure~\ref{fig_map_hcn_3_0__4_3__5_8_K_0}  shows $\rho_{ J K_J=0 }$  for $J^{\pi} = 3_1^-, 4_1^+, 5_1^-$ resonances, which are expected -- based on energy considerations -- to form a continuation of the ground state rotational band. One can see that these densities are not only drastically different from those of ${0}^{+}_1$, ${1}^{-}_1$, and ${2}^{+}_1 $ states but also change from one state to another. It is also worth noting that the densities of $3_1^-, 4_1^+$, and $5_1^-$ resonances have spatial extensions  that are dramatically  larger as compared to the three bound members of the ground-state band. 

\begin{figure}[htb]
	\includegraphics[width=0.8\columnwidth]{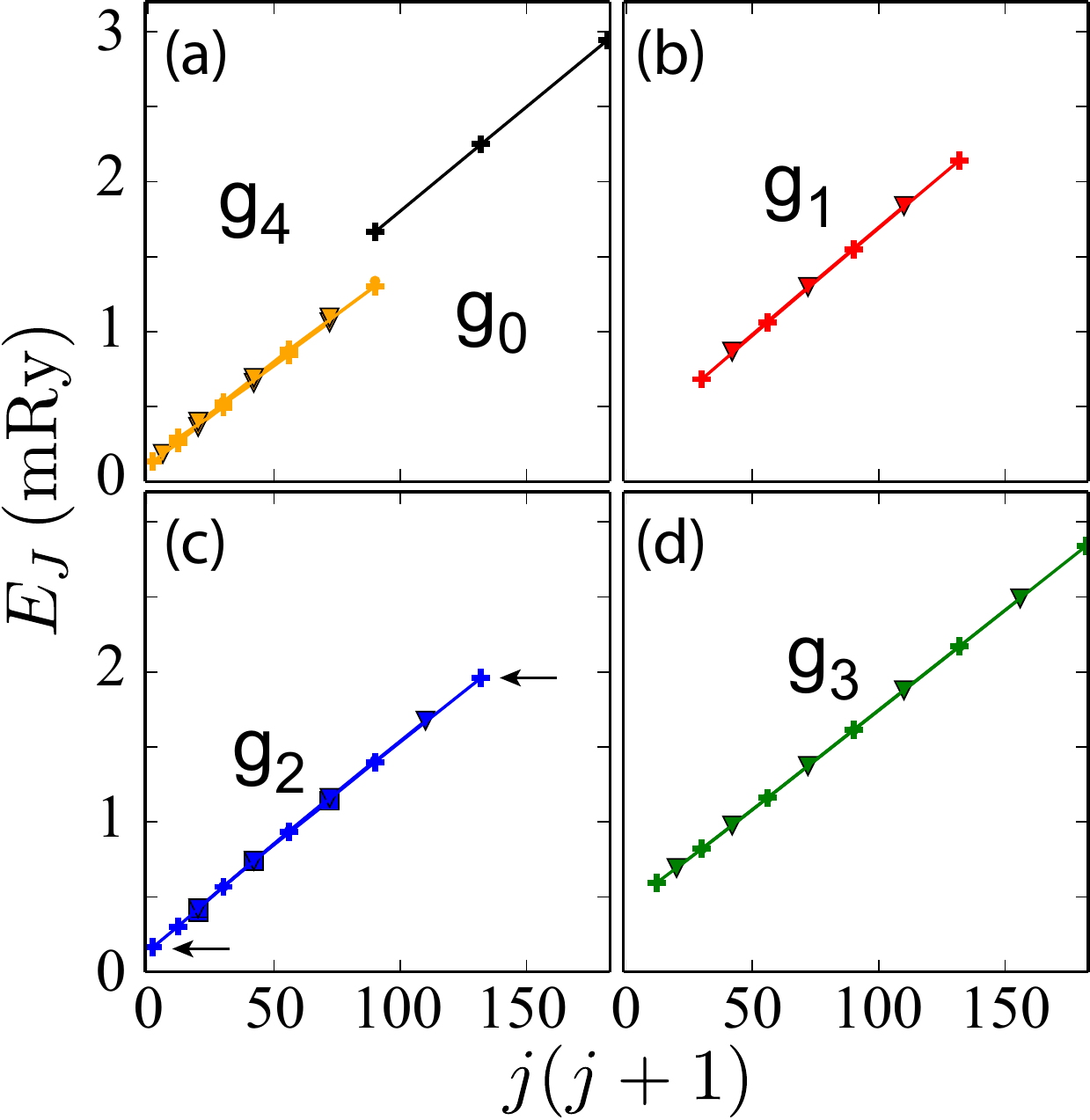}
	\caption{(Color online) Excitation energies of resonances of the HCN${ {}^{-} }$ dipolar anion for various $J^\pi$  as a function of ${ j ( j + 1 ) }$, 
	where $j$ is the rotational angular momentum of the molecule in the dominant channel wave function for each considered state. 
	Colors are related to groups of states in the complex-energy plane identified in Fig.~\ref{fig_spectra}. The symbols 
	$ \blacksquare $, $ \bullet $, $ \blacktriangledown $ and $ + $ denote states with ${ J^\pi = 2^+ }$, ${ 3^- }$, ${ 4^+}$ and ${ 5^- }$, respectively.}
	\label{fig_plot_bands_j_sep}
\end{figure}
\begin{figure}[htb]
	\includegraphics[width=0.8\columnwidth]{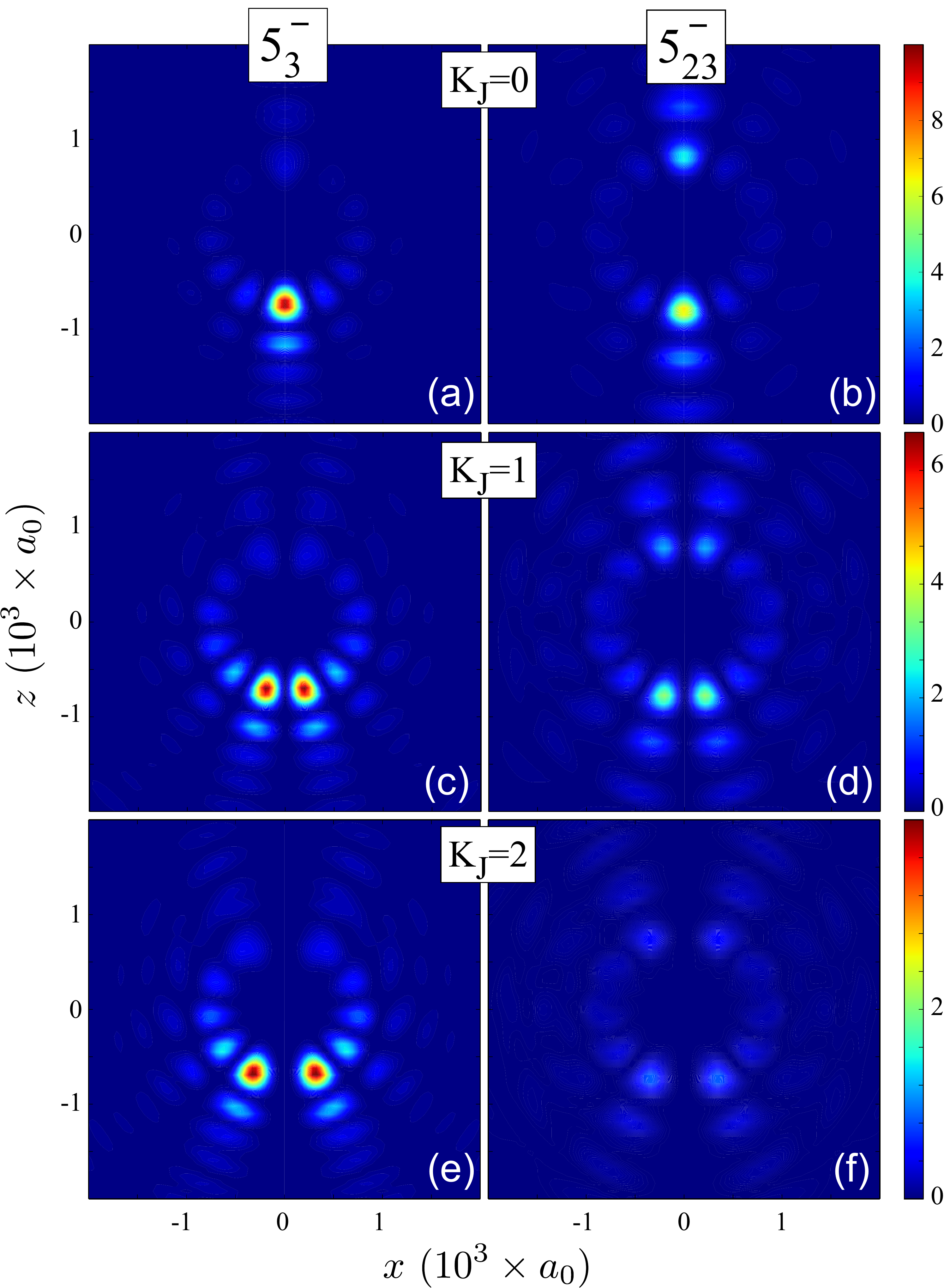}
	\caption{(Color online)  Intrinsic densities $\rho_{ J K_J } (r,\theta)$ (in $10^{-10}\,a_0$) with 
	$K_J=0, 1, 2$, for the two  resonances $5^-_3$ and $5^-_{23}$ belonging to the group $g_2$, marked by arrows in 
	Fig.~\ref{fig_plot_bands_j_sep}(c). For both states, the dominant channel has $\ell=6$.}
	\label{fig_map_hcn_5_24-415}
\end{figure}

As seen in Fig.~\ref{fig_spectra}, there appear clusters of resonances having the same total angular momentum ${ J }$ within one group $g_i$.
In each cluster, dominant channel wave functions have the same orbital angular momentum of the valence electron ${\ell}$, but different rotational angular momenta of the molecule ${ j }$. Excitation energies of resonances are plotted as a function of the molecular angular momentum ${ {j} }$ in 
Fig.~\ref{fig_plot_bands_j_sep} for different groups of resonances of Fig.~\ref{fig_spectra}. It is seen that these states form very regular rotational band sequences in $j$ rather than in $J$. Different members of such bands lie close in the complex energy plane and have similar densities 
$\rho_{ J K_J } (r,\theta)$. This is illustrated in Fig.~\ref{fig_map_hcn_5_24-415}, which shows $\rho_{ J K_J } (r,\theta)$ for
the two $J^\pi=5^-$ resonances marked by arrows in Fig.~\ref{fig_plot_bands_j_sep}(c); namely $5^-_3$, having the dominant parentage $(\ell,j)=(6,1)$,
and $5^-_{23}$, having the dominant parentage $(6,11)$.  

\begin{figure}[htb]
\includegraphics[width=0.8\columnwidth]{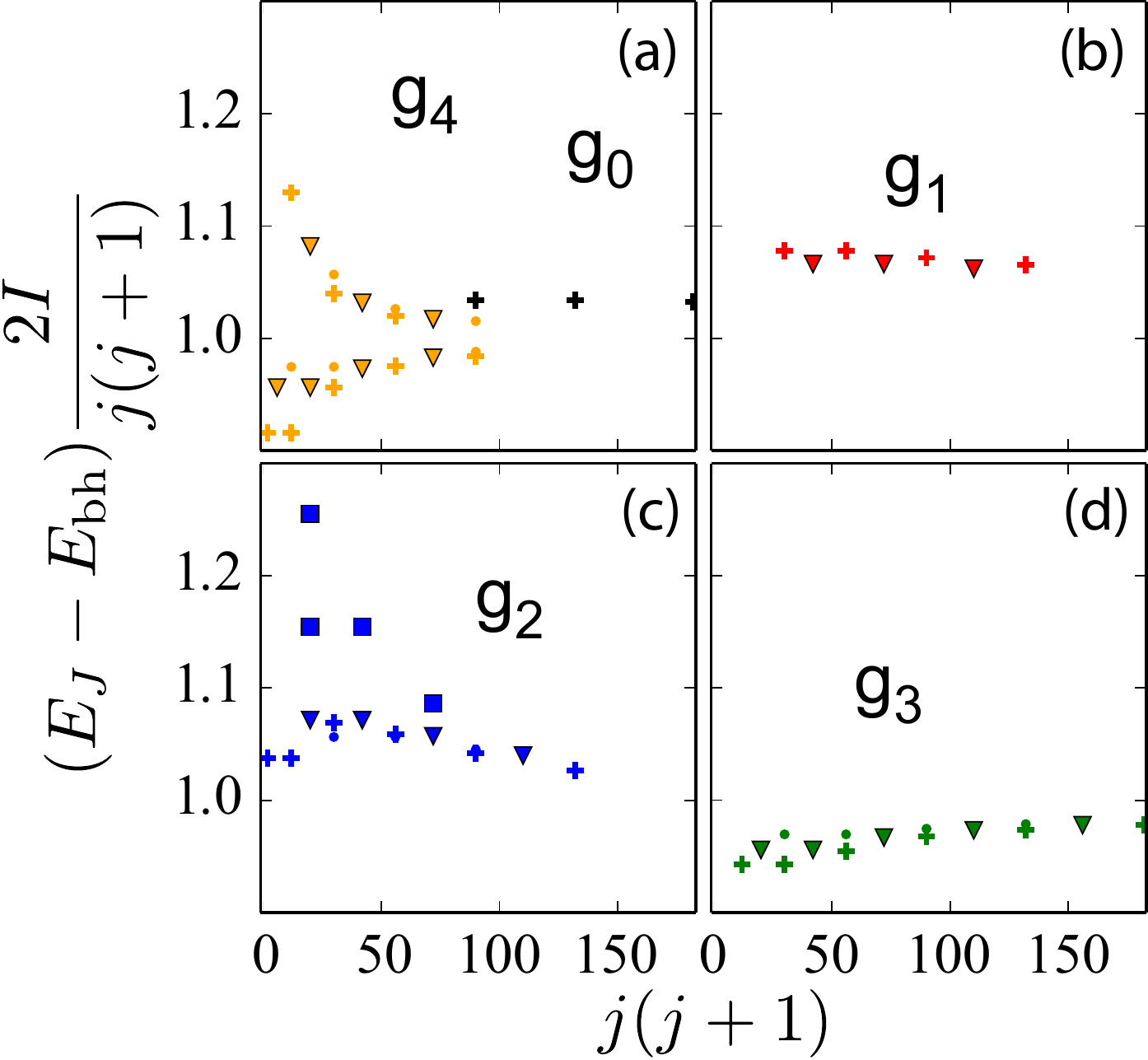}
	\caption{(Color online) Similar as in Fig.~\ref{fig_plot_bands_j_sep} but 
	for $(E_J-E_{\rm bh})\frac{2I}{j(j+1)}$, where $E_{\rm bh}$ is a bandhead energy at $j=0$.}
	\label{fig_plot_bands_j_rescaled_sep}
\end{figure}

The results of Fig.~\ref{fig_plot_bands_j_sep}  suggest that the rotational resonance structures are governed by a weak $\ell$-$j$ coupling, whereby the orbital motion of a valence electron is decoupled from the rotational motion of a dipolar neutral molecule. To illustrate the weak coupling better,  
in Fig.~\ref{fig_plot_bands_j_rescaled_sep} we display  the rotational bands of Fig.~\ref{fig_plot_bands_j_sep}  with respect to the rigid rotor reference
$j(j+1)/2I$.
In the  case of a perfect $\ell$-$j$ decoupling, the rescaled energy in Fig.~\ref{fig_plot_bands_j_rescaled_sep} should be equal to 1. One can  see that this limit is reached in most cases, with deviations from unity being  less than 10 \%. Larger deviations are found for few low-$j$ states in bands with $ J=2 $ in $g_2$ and $ J=5 $ in $g_4$. Consequently, intrinsic densities for resonances in these two bands exhibit certain differences, whereas they are almost identical for bands close to the weak-coupling limit. 

The variations seen in  Fig.~\ref{fig_plot_bands_j_rescaled_sep} can be traced back to the  leading channel components along a $j$-band. 
\begin{table}[htb]	
\caption{\label{tab_mean_val_norm}
Contributions of the two leading channel wave functions to the norm of resonances in different groups of states in Fig.~\ref{fig_spectra}. Only states with dominant channel $\ell= 6 $ for $g_1$, $g_2$, annd $g_4$, and $\ell= 8$ for $g_0$, and $g_3$ are included.}
			\begin{ruledtabular}
	\begin{tabular}{ccccc}
		Group 	& \multicolumn{4}{c}{ $\ell$ of dominant channels } \\
		& 6 & 7 & 8 & 9 \\
		\hline \\[-6pt]
		$g_0$ 	& -- 	& --	& 60\% 	& 40\% \\
		$g_3$ 	& 1\% 	& --	& 99\% 	& -- \\
		\hline \\[-6pt]
		$g_1$	& 70\% 	& 30\% 	& -- 	& -- \\
		$g_2$ 	& 90\% 	& 10\% 	& -- 	& -- \\
		$g_4$ 	& 100\% & -- 	& --	& -- 
	\end{tabular}
\end{ruledtabular}
\end{table}
Table~\ref{tab_mean_val_norm} displays the leading channel wave functions to the resonances in different groups $g_i$. Not surprisingly, the resonances forming $j$-band structures are associated with high orbital angular momentum components $\ell=6-9$ for which the centrifugal force induces a strong decoupling of the electron and the rotor. For regular bands in 
Fig.~\ref{fig_plot_bands_j_rescaled_sep}, the $\ell$-content  is almost constant as a function of $j$. For instance, for the four $J=5$ states in $g_1$, the $(\ell,j)$ parentages of the two largest (6,$j$)/(7,$j+1$) components are: 0.64/0.37 ($j=5$), 0.67/0.36 ($j=7$), 0.69/0.35 ($j=9$), and 0.70/0.34 ($j=11$). On the other hand, for bands that exhibit stronger $j$-dependence in Fig.~\ref{fig_plot_bands_j_rescaled_sep} the $\ell$-compositions change.  

Interesting complementary information about the arrangement of resonances in the continuum of HCN$^-$ 
\begin{figure}[htb]
\includegraphics[width=0.80\columnwidth]{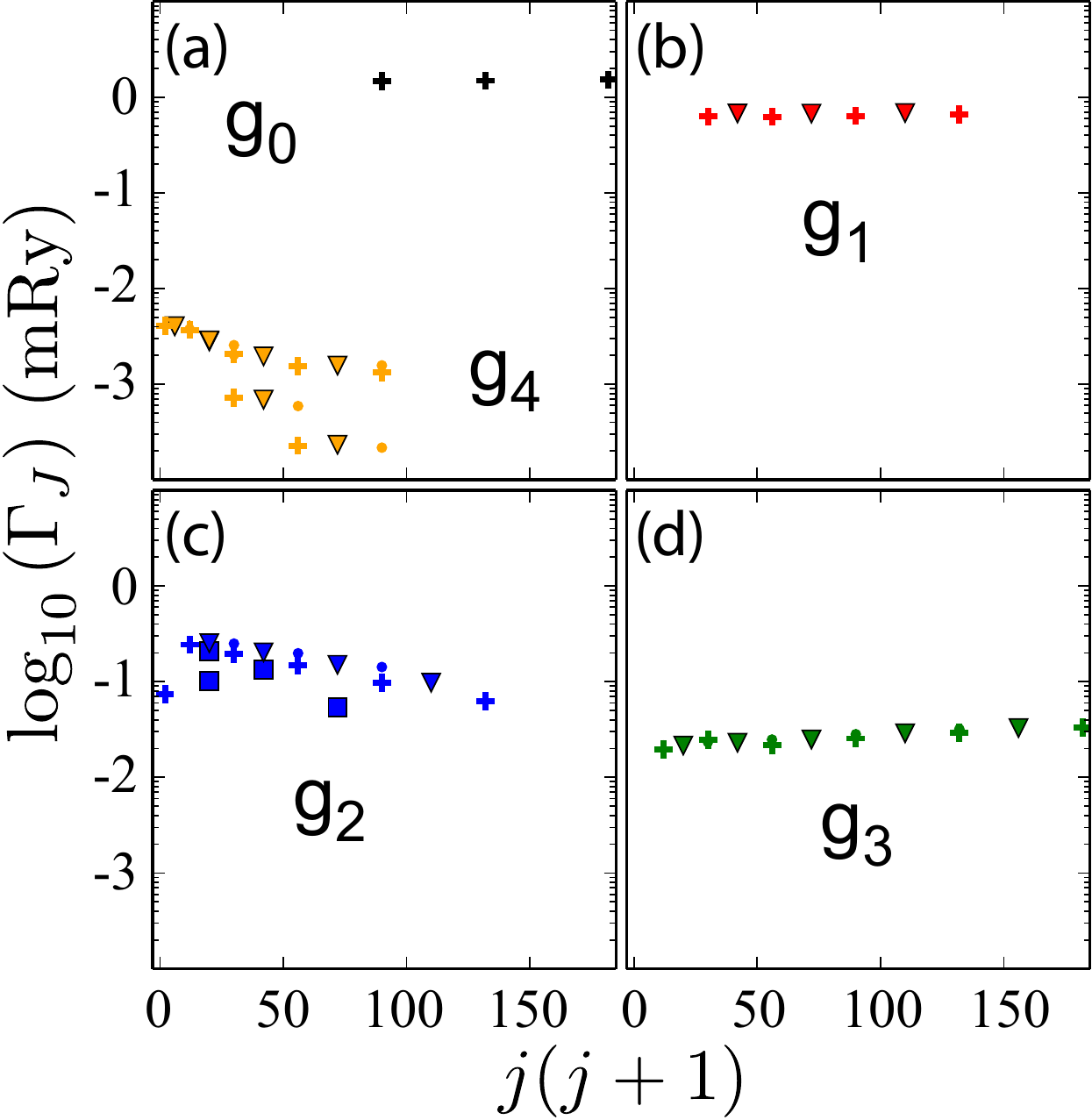}
	\caption{(Color online) Similar as in  Fig.~\ref{fig_plot_bands_j_sep} but for the resonance widths. }
	\label{fig_plot_width_j}
\end{figure}
can be seen in Fig.~\ref{fig_plot_width_j} which shows the decay width for various $j$-bands in different groups $g_i$ and different total angular momenta $J$ within a given group. One can see that the  bands that exhibit largest deviations from the weak-coupling limit in Fig.~\ref{fig_plot_bands_j_rescaled_sep},   also show strong in-band variations of the decay width. In regular bands belonging to $g_0$, $g_1$, and $g_3$, the width stays constant or slightly increases with  $j$. On the other hand, the irregular bands in $g_2$ and $g_4$ exhibit a decrease of $\Gamma_J$  with  $j$.  Such a behavior of lifetimes can be traced back to variations of the $(\ell, j)$-content of the resonance wave function with rotation.

\section{Conclusions}\label{conclusions}

In this work, we studied bound and resonance states of the dipole-bound anion of hydrogen cyanide HCN$^-$ using the  open-system Berggren expansion method. To identify the decaying resonant states and separate them from the scattering background, we adopted  the algorithm based on contour shift in the complex energy plane.  To characterize spatial distributions of valence electrons, we introduced the intrinsic density of the valence electron. This quantity is useful when assigning resonant states into rotational bands.

Non-adiabatic coupled-channel calculations with a  pseudo potential adjusted to ground-state properties of HCN$^-$ predict only three bound states of the dipole-bound anion: $0^+$, $1^-$, and $2^+$. Those states are members of the ground-state rotational band. The lowest $3^-_1$ state is a threshold  
resonance; its intrinsic structure is very different from that of $0^+_1$, $1^-_1$, and $2^+_1$ states, and the lowest-energy resonances $4^+_1$, 
and $5^-_1$.  

The dissociation threshold in the HCN${ {}^{-} }$ dipolar anion defines two distinct regimes of rotational motion. Below the threshold, rotational bands in $J$ can be associated with bound states. Here, the valence electron follows the collective rotation of the molecule. This is not the case above the threshold where the motion of a valence electron in a resonance state is largely decoupled from the molecular rotation with the families of resonances forming regular band sequences in $j$. Widths of resonances forming  $j$-bands depend primarily on the electron's orbital angular momentum in the dominant channel and remain fairly constant within each band for regular bands. Small irregularities in  moments of inertia and decay width are predicted for   very narrow resonances in the vicinity of the dissociation threshold.

In summary, this work demonstrates the feasibility of accurate calculations of weakly bound and unbound states of the dipolar anions using the Berggren expansion approach. Our prediction of two distinct modes of rotation in this open quantum system awaits experimental confirmation. It is interesting to note a similarity between the problem of a dipolar anion and  a coupling of  electrons in high molecular Rydberg states to molecular rotations \cite{Dagata89,Remacle96}. Namely, in both cases one deals with non-adiabatic coupling of a slow electron to  the fast rotational motion of the core, with no  separation in the single-particle and collective time scales.

\begin{acknowledgments}
Discussions with  R.N. Compton and W.R. Garrett are gratefully acknowledged.
This material is based upon work supported by the U.S. Department of Energy, Office of Science, Office of Nuclear Physics under Award Number No.\ DEFG02-96ER40963 (University of Tennessee). This work was supported partially through FUSTIPEN (French-U.S. Theory Institute for Physics with Exotic Nuclei) under DOE grant number DE-FG02-10ER41700.
\end{acknowledgments}

\bibliographystyle{apsrev4-1}
\bibliography{ANIONS}

\end{document}